\documentclass[aps,pra,amsmath,amssymb,floatfix,twocolumn,amsmath,superscriptaddress,twocolumn,nofootinbib,tighten,letterpaper]{revtex4-2}
\usepackage[colorlinks,linkcolor=blue,citecolor=blue,urlcolor=blue]{hyperref}
\usepackage{multirow}
\usepackage{subfigure}
\usepackage{color}
\usepackage{mathrsfs}
\usepackage{hyperref}
\usepackage[normalem]{ulem}
\usepackage{bm}

\usepackage{amssymb}
\usepackage{amsmath}
\renewcommand\vec[1]{\ensuremath\boldsymbol{#1}}

\usepackage{tabularray}

\usepackage{array} 
\newcolumntype{P}[1]{>{\centering\arraybackslash}p{#1}}

\definecolor{RowColor}{rgb}{0.88,1,0.9}

\usepackage{amsfonts, relsize, color, mathtools, physics}
\usepackage{graphics}
\usepackage{graphicx}
\usepackage{subfigure}
\usepackage{color}
\usepackage{comment}

\begin{document}

\title{Upper critical dimension for dirty Weyl semimetal-to-metal quantum phase transitions}

\author{Yongtai Li}
\affiliation{Department of Physics, Lehigh University, Bethlehem, Pennsylvania 18015, USA}

\author{Bitan Roy}
\affiliation{Department of Physics, Lehigh University, Bethlehem, Pennsylvania 18015, USA}

\date{\today}

\begin{abstract}
Weyl or Dirac fermions with the iconic linear energy-momentum relation and average density of states (ADOS) $\rho(E) \sim |E|^{d-1}$ at energy $E$ in $d$ spatial dimensions, constitute a unique setup to study the disorder-driven semimetal-to-metal quantum phase transition (QPT) between ballistic (realized for weak disorder) and diffusive (stabilized at stronger disorder) quasiparticles. Such a QPT takes place only for $d>2$ and falls beyond the realm of the Anderson metal-to-insulator transition. From numerically computed ADOS (using the kernel polynomial method) in dirty Weyl systems in $d=2$ to $6$, here we show that $d=2$ and $d=4$ are the lower ($d_\ell$) and upper ($d_u$) critical dimensions for such an unconventional QPT, respectively, as suggested from the solution of quasiparticle lifetime within the self-consistent Born approximation. Consequently, for $d \geq 4$ the associated correlation length exponent is found to be $\nu \approx 0.5$ (within numerical accuracy). However, the dynamic scaling exponent at the quantum critical point is pinned close to $z \approx d/2$ (numerically) for any $d \geq 3$, which is shown to be an exact result from a field-theoretic renormalization group calculation. Therefore, Weyl semimetal-to-metal QPTs can be studied field theoretically around both $d_\ell$ and $d_u$.       
\end{abstract}

\maketitle

\emph{Introduction}.~Transitions between two distinct phases of matter, tuned by nonthermal parameters, give rise to the venerable concept of quantum phase transition (QPT) that strictly takes place at zero temperature where all the thermal agitation freezes. In quantum materials, the interaction-to-bandwidth ratio tunable via hydrostatic pressure, for example, impurity concentration or disorder, magnetic field etc.\ serve as experimentally tunable nonthermal parameters to trigger a plethora of QPTs. In this context, impurity concentration or strength of disorder is unique as it allows the possibility of Anderson metal-to-insulator transition (AMIT) in noninteracting systems. Such a QPT takes place only in dimensions $d>2$ with the lower critical dimension (LCD) $d_\ell=2$~\cite{Anderson1958, Abrahams1979, Wegner1976, Wegner1980, Vollhardt1982, Schreiber1996, Lee1985}. It is also conceivable for a QPT to be characterized by an upper critical dimension (UCD) $d_u$, at and above which the associated quantum critical phenomena are mean-field or Gaussian in nature. Specifically, at $d=d_u$ violation of the hyperscaling hypothesis leads to logarithmic corrections to scaling of physical observables~\cite{ZinnJustin2002, Herbut2007, Sachdev2011}. However, the AMIT is devoid of any finite $d_u$, for which $d_u \to \infty$~\cite{Harris1981, Mirlin1994, Castellani1986, Garcia2007, Mard2017}.

\begin{figure}[t!]
    \centering
    \includegraphics[width=0.95\linewidth]{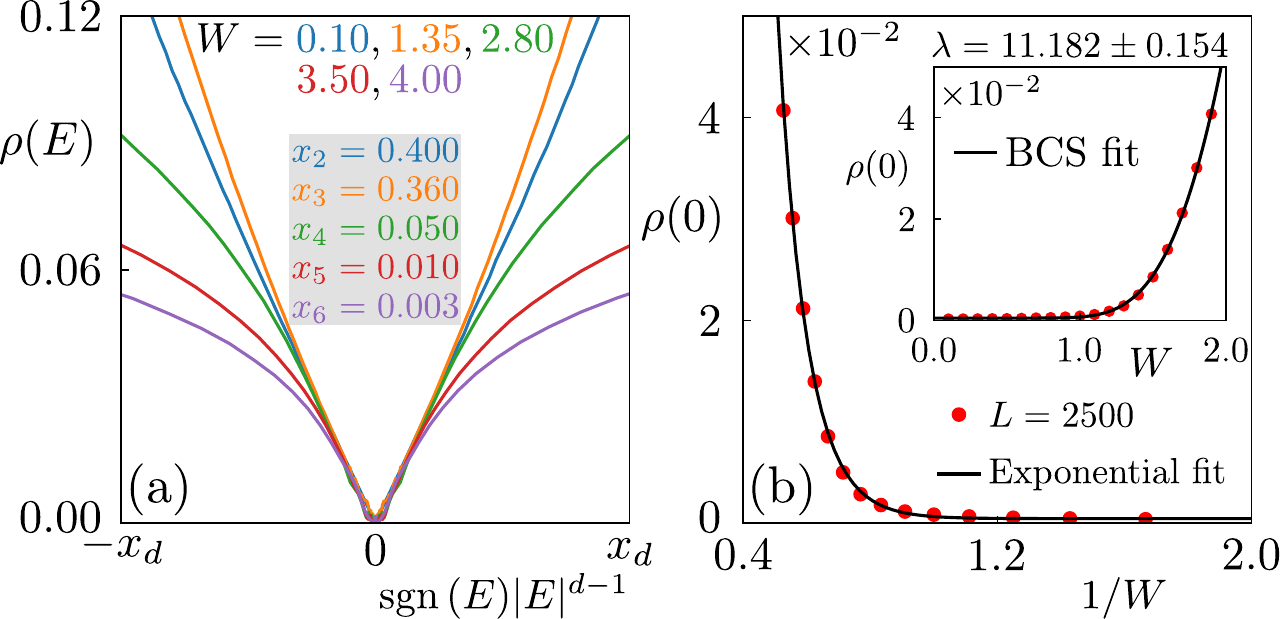}
    \caption{(a) ADOS $\rho(E)$ at energy $E$ with ${\rm sgn}(E) \; |E|^{d-1}$ in $d$-dimensional WSMs at weak disorder $W$ (see legend). (b) Scaling of $\rho(0)$ with $1/W$ for a WSM in $d=2$. Inset: Scaling of $\rho(0)$ with $W$. Red dots are numerically obtained data points and black curves are the fits with $\rho(0) \sim \exp(-\lambda/W)$, where $\lambda$ is a fitting parameter (see legends).    
    }
    \label{fig:clean_2DBCS}
\end{figure}

In this respect, Weyl semimetals (WSMs) displaying a linear energy-momentum relation that leads to $\rho(E) \sim |E|^{d-1}$ scaling of the average density of states (ADOS) at energy $E$ [Fig.~\ref{fig:clean_2DBCS}(a)] are unique in the presence of disorder. Namely, ballistic quasiparticles in weakly disordered WSMs are stable in any $d>2$ and at a moderate disorder the system undergoes a continuous QPT into a compressive diffusive metal (CDM)~\cite{Fradkin1986, Murakami2009, goswami2011, Imura2013, Herbut2014, roydassarma2014, Moon2014, Syzranov2015a, Syzranov2015b, Altland2015, pix2015, Syzranov2016, Pix2016a, roydassarma2016, bera2016, Gorbar2016, RoyJuricicDasSarma2016, RoyDasSarma2016PRB, Carpentier2016, Carpentier2017, Justin2017, Goswami2017a, Goswami2017b, Alavirad2017, Slager2017, WilsonRefael2018, Slager2018, Carpentier2018, Carpentier2019, Szabo2020, Gruzberg2024, Tyner2025}. Such a conclusion holds up to nonperturbative rare region effects that yield a CDM with an exponentially small $\rho(0)$ possibly for infinitesimal disorder~\cite{Huse2014, Huse2016a, Huse2016b, buch2018, JustinWilson2020, JustinWilson2024, PixleyWilsonReview2021}, however, without altering the critical properties. Hence, the WSM-to-metal QPT is characterized by an LCD of $d_\ell=2$, similar to the AMIT. However, in stark contrast to the AMIT, the WSM-to-metal QPT is believed to possess a UCD of $d_u=4$, which we establish here from the following key observations, obtained by combining extensive numerical and analytical calculations. Identical conclusions hold for Dirac semimetals (not shown explicitly).

\begin{figure*}[t!]
    \centering
    \includegraphics[width=0.90\linewidth]{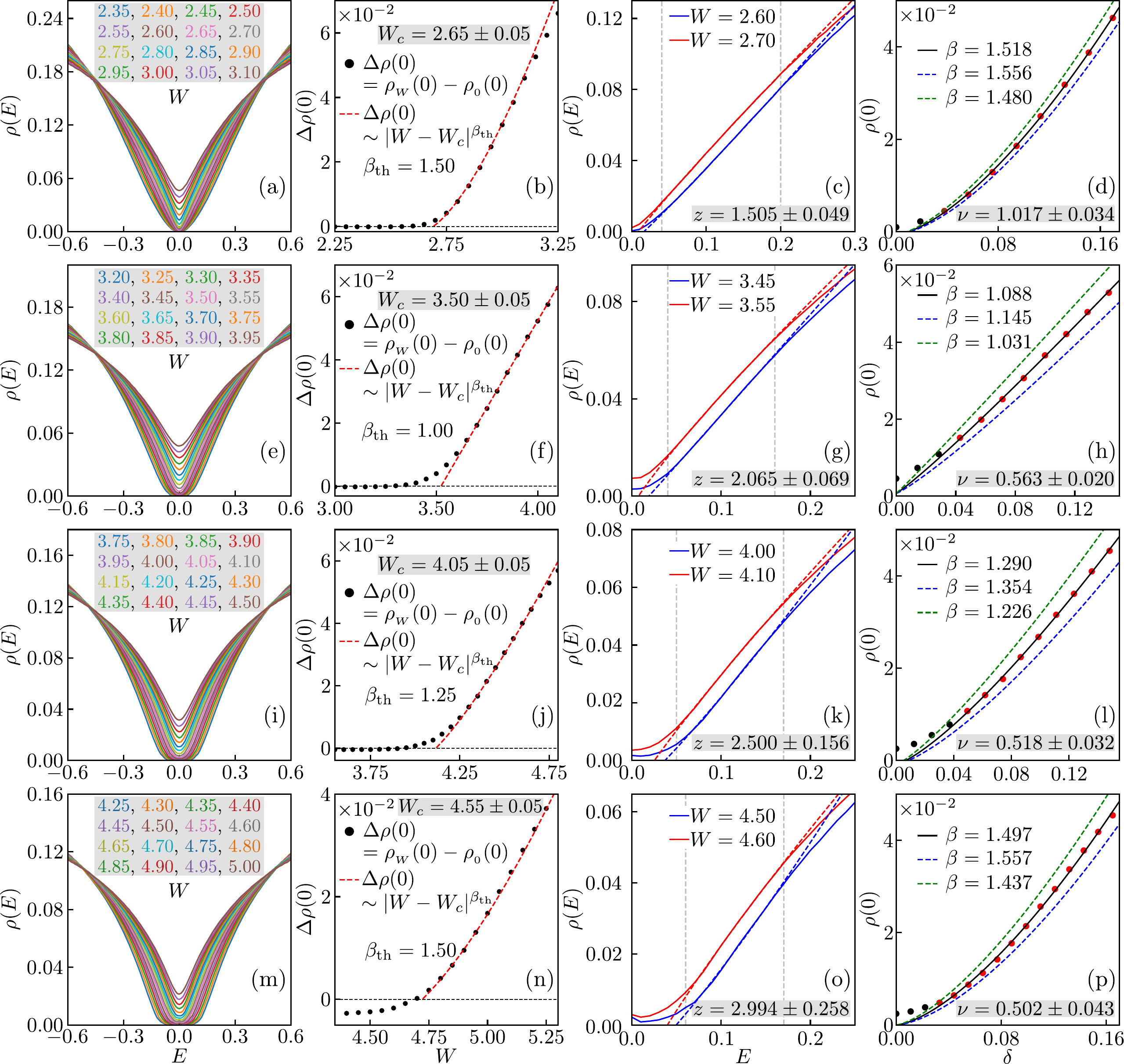}
    \caption{(a) ADOS $\rho(E)$ at energy $E$ as a function of disorder $W$. (b) $\Delta \rho(0)$ (see text) as a function of $W$ (black dots), marking the critical disorder strength $W_c$, with the red dashed line showing its expected behavior from analytical calculations. (c) DSE $z$ near $W=W_c$, obtained from the power-law fitting of $\rho(E)$ within the range of $E$ marked by vertical dashed lines. (d) CLE $\nu$, obtained from the scaling of $\rho(0)$ in the metallic phase from red (by excluding black) colored data points. All the calculations are performed in the largest system with $L=200$ in $d=3$. Panels (e)-(h) are analogous to panels (a)-(d), respectively, but in the largest system with $L=52$ in $d=4$. Panels (i)-(l) are analogous to panels (a)-(d), respectively, but in the largest system with $L=24$ in $d=5$. Panels (m)-(p) are analogous to panels (a)-(d), respectively, but in the largest system with $L=12$ in $d=6$. See the legends for details and values of $W_c$, $z$, and $\nu$.  
    }
    \label{fig:dataanalysis}
\end{figure*}

\begin{figure}[t!]
    \centering
    \includegraphics[width=0.95\linewidth]{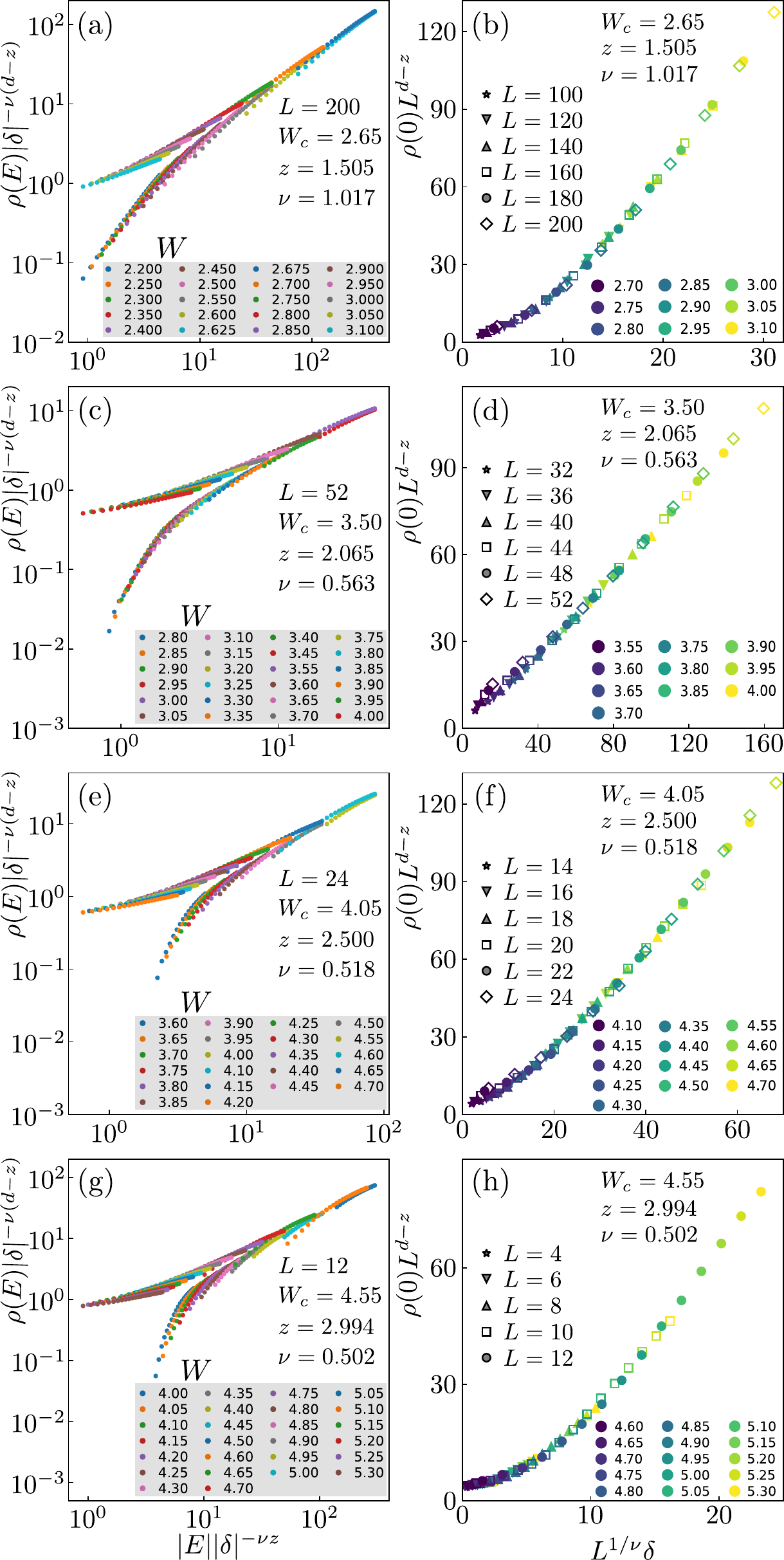}
    \caption{(a) Data collapse in the largest system in $d=3$, obtained by comparing $\rho(E) |\delta|^{-\nu (d-z)}$ with $|E|\delta^{-\nu z}$ with numerically obtained mean values of $W_c$, $z$, and $\nu$ (see legend). (b) Data collapse in the metallic phase, obtained by comparing  $\rho(0) L^{d-z}$ with $L^{1/\nu} \delta$ in $d=3$ for various system sizes $L$ (see legend). Panel (c) [(d)] is analogous to (a) [(b)], but in $d=4$. Panels (e) [(f)] is analogous to (a) [(b)], but in $d=5$. Panel (g) [(h)] is analogous to (a) [(b)], but in $d=6$.}
    \label{fig:datacollapse}
\end{figure}

\emph{Key results}.~Numerically computed ADOS in $d$-dimensional dirty WSMs with $d=2, \cdots, 6$ here, obtained by employing the kernel polynomial method (KPM) lies at the heart of the present analysis~\cite{Weisse2006, Groth2014}. From the scaling of $\rho(0)$, we show that $d=2$ is the LCD for the WSM-to-metal QPT. The system then becomes a \emph{putative} CDM (ultimately becoming an Anderson insulator in the thermodynamic limit) for infinitesimal disorder $W$ [Fig.~\ref{fig:clean_2DBCS}(b)]. By contrast, in any $d \geq 3$, WSMs are stable against sufficiently weak disorder, where $\rho(0) =0$, and undergo a QPT into a CDM at finite strength of disorder ($W_c$) with the associated UCD $d_u=4$. Such outcomes are anticipated from the solution for the quasiparticle lifetime ($\tau_{\rm qp}$), obtained within the self-consistent Born approximation (SCBA) as $\tau^{-1}_{\rm qp} \propto \rho (0)$. Within numerical accuracy, we find that the correlation length exponent (CLE) $\nu \approx 0.5$ for any $d \geq 4$, indicating a Gaussian or mean-field nature of WSM-to-metal QPTs therein. We find that the dynamic scaling exponent (DSE) $z$ near the WSM-to-metal quantum critical point (QCP) is $z \approx d/2$, leading to an $E$-linear scaling of the ADOS in any $d \geq 3$. See Fig.~\ref{fig:dataanalysis}. Such an outcome is shown to be exact from a renormalization group (RG) calculation. Numerically obtained $W_c$, $z$, and $\nu$ produce excellent data collapses [Fig.~\ref{fig:datacollapse}].

\emph{Anticipation from SCBA}.~The stability of Weyl fermions against disorder, the possibility of WSM-to-metal QPTs, and the existence of associated LCD and UCD can be anticipated from the solution of $\tau^{-1}_{\rm qp}$ which within the SCBA is obtained from (for $\hbar=1$)~\cite{Ping2006} 
\begin{equation}~\label{eq:SCBA}
W \int^{E_\Lambda}_0 \; \frac{\rho(E)}{E^2 + \tau^{-2}_{\rm qp}} \; dE=1,  
\end{equation}
where $E_\Lambda$ is an ultraviolet energy cutoff up to which $\rho (E) \sim |E|^{d-1}$. In $d=2$, as $E_\Lambda \to \infty$ the above integral shows a logarithmic divergence, yielding $\tau^{-1}_{\rm qp} \sim \exp[-1/W]$ for $\tau^{-1}_{\rm qp} E_\Lambda \ll 1$, indicating the onset of a metallic phase for infinitesimal disorder with $\rho(0)$ displaying a BCS-like scaling with $W$. Thus, $d=2$ is the LCD associated with the WSM-to-metal QPT. However, for  any $d>2$, the above integral displays a \emph{power-law} ultraviolet divergence as $E_\Lambda \to \infty$, suggesting that $\tau^{-1}_{\rm qp}$ becomes finite only for $W>W_c$, where $W_c=\int^{E_\Lambda}_0 dE \rho(E)/E^2$. After taking $E_\Lambda \to \infty$, we find $\tau^{-1}_{\rm qp}= 2 \bar{\delta}/\pi$ in $d=3$, where $\bar{\delta}=W^{-1}_c-W^{-1}$. By contrast, in $d=4$ we obtain $\bar{\delta}=\tau^{-2}_{\rm qp} \; \ln\left( [\tau^{-2}_{\rm qp}+E^2_\Lambda]/\tau^{-2}_{\rm qp}\right)/2$. The logarithmic correction to the finite self-consistent solution of $\tau^{-1}_{\rm qp}$, which can only be found for $\bar{\delta}>0$, captures the violation of hyperscaling and suggests that $d=4$ is the UCD associated with the WSM-to-metal QPT. Next we proceed to substantiate these expectations from numerically obtained ADOS in dirty WSMs in $d=2, \cdots, 6$.

\emph{Lattice model}.~We consider the Bloch Hamiltonian on a $d$-dimensional hypercubic lattice with lattice spacing $a$ in all directions that supports linearly dispersing gapless Weyl quasiparticles around $2^d$ number of high-symmetry points of the corresponding Brillouin zone 
\begin{equation}
H^{\rm Bloch}_{\rm Weyl} (\vec{k})= t \sum^{d}_{j=1} \; \sin(k_j a) \; \Gamma_j,  
\end{equation}
where $t$ is the nearest-neighbor hopping amplitude, $k_j$ is the component of spatial momentum $\vec{k}=(k_1, \cdots, k_d)$ in the $j$th direction, and $\Gamma_j$s are mutually anticommuting Hermitian matrices, satisfying the Clifford algebra $\{ \Gamma_j, \Gamma_k \}= 2 \delta_{jk}$ (Kronecker delta symbol). The particle-hole symmetric energy spectrum of $H^{\rm Bloch}_{\rm Weyl} (\vec{k})$ is $\pm E_{\vec{k}}$ with $E_{\vec{k}}=t \left[ \sin^2(k_1 a) + \cdots + \sin^2(k_d a)\right]^{1/2}$. Around the band-touching Weyl points, we recover linear energy-momentum relation $\pm E_{\vec{q}}$ with $E_{\vec{q}}= v_{_{\rm F}}|\vec{q}|$, where $\vec{q}$ is the small momentum measured from such diabolic points.

The dimensionality of the $\Gamma$ matrices ($d_\Gamma$) depends on $d$. In any $d$, we need exactly $d$ mutually anticommuting $\Gamma$ matrices. We choose $\Gamma_j \equiv \sigma_j$ for $j=1,2,3$ in $d=2$ and $3$ with $d_\Gamma=2$. For $d=4$ and $d=5$, we work with the representation $\Gamma_j=\tau_3 \sigma_j$ for $j=1,2,3$, $\Gamma_4=\tau_1 \sigma_0$, and $\Gamma_5=\tau_2 \sigma_0$ with $d_\Gamma=4$. Finally, as in $d=6$ $H^{\rm Bloch}_{\rm Weyl} (\vec{k})$ involves six such $\Gamma$ matrices, they belong to the representation $\Gamma_j=\eta_3 \tau_3 \sigma_j$ for $j=1,2,3$, $\Gamma_4=\eta_3 \tau_1 \sigma_0$, $\Gamma_5=\eta_3 \tau_2 \sigma_0$, and $\Gamma_6=\eta_1 \tau_0 \sigma_0$ with $d_\Gamma=8$. Here $\sigma_j$, $\tau_j$, and $\eta_j$ are two sets of Pauli matrices with $j=1,2,3$, and $\sigma_0$, $\tau_0$, and $\eta_0$ are identity matrices. However, our conclusions are insensitive to $\Gamma$ matrix representation.

To compute ADOS using KPM in disordered WSMs, we consider the corresponding tight-binding Hamiltonian
\begin{equation}
H^{\rm TB}_{\rm Weyl}= \frac{t}{2i}  \sum_{\vec{r}} \sum^{d}_{j=1} \Psi^\dagger_{\vec{r}} \Gamma_j \Psi_{\vec{r}+\hat{e}_j} 
+ \sum_{\vec{r}} V(\vec{r}) \Psi^\dagger_{\vec{r}} \Gamma_0 \Psi_{\vec{r}},
\end{equation}
where $\Gamma_0$ is a $d_\Gamma$-dimensional identity matrix, $\Psi^\top_{\vec{r}}$ is a row vector containing fermionic annihilation operators at $\vec{r}$, involving the internal degrees of freedom, acting on the space of the $\Gamma$ matrices, and $\hat{e}_j$ is the unit vector in the $j$th direction. On each lattice site, we sample pointlike charge impurities $V(\vec{r})$ randomly and uniformly from a box distribution $[-W/2, W/2]$, where the box width $W$ is the disorder strength. Throughout, we impose periodic boundary conditions in all directions.

\emph{KPM and ADOS}.~We employ KPM to compute the disorder-averaged, denoted by $\langle \cdots \rangle$, ADOS at energy $E$
\begin{equation}
\rho(E) = \biggl< \; \frac{1}{L^d \; d_\Gamma} \sum^{L^d}_{j=1} \delta(E-E_j) \; \biggr>,
\end{equation}
where $L$ is the linear dimension of the hypercubic lattice in each direction. Even though ADOS is a self-averaging quantity, we still average it over four independent disorder realizations to minimize the residual statistical error. We take stochastic trace over 12 unimodular random vectors and typically compute $N_m=4096$ Chebyshev moments while extracting $\rho(E)$~\cite{Weisse2006}. The WSM and CDM phases are characterized by $\rho(0) \to 0$ and a finite $\rho(0)$, respectively. Hence, ADOS at zero energy serves as an order parameter across the QPT between them. The largest system sizes in this work are $L_{\rm max}=2500, 200, 52, 24$, and $12$ in $d=2,3,4,5$, and $6$, respectively, with comparable total number of sites.

In our numerical calculations, as $d$ increases, $L_{\rm max}$ has to be reduced due to limited computational resources, which introduces a finite ADOS close to $E=0$ even in clean systems (finite-size effects). Furthermore, momentum quantization with periodic boundary conditions introduces a finite number of exact zero-energy modes, which nonetheless shift away from $E=0$ in the presence of disorder as they break translational invariance. Thus, instead of tracking $\rho(0)$ as a function of $W$ to locate the WSM-to-insulator QCP at $W=W_c$ where it becomes finite, we focus on $\Delta \rho(0)= \rho_{_W}(0)-\rho_{_0}(0)$, where $\rho_{_0}(0)$ [$\rho_{_W}(0)$] is the ADOS at $E=0$ in the clean system [for disorder strength $W$]. Then, $\Delta \rho(0)<0$ for $W<W_c$, $\Delta \rho(0)>0$ for $W>W_c$, and $\Delta \rho(0)=0$ at $W=W_c$~\cite{Tyner2025}. Irrespective of whether we track $\rho(0)$ or $\Delta \rho(0)$, we find almost the same values for $W_c$ in $d=3,4$, and $5$. But, in $d=6$ due to small $L_{\rm max}$, the scaling of $\Delta \rho(0)$ with $W$ gives more reliable estimation of $W_c$.

\emph{Scaling}.~ADOS follows the scaling form~\cite{Herbut2014, Pix2016a, bera2016, Slager2018}
\allowdisplaybreaks[4]
\begin{equation}~\label{eq:ADOSscaling}
\rho(E)= \delta^{(d-z)\nu} \; {\mathcal G} \big( |E| \delta^{-\nu z}, L^{1/\nu} \delta \big),
\end{equation}
where $\delta=(W-W_c)/W_c$ is the reduced distance from the WSM-to-metal QCP and ${\mathcal G}$ is an otherwise unknown universal function of its arguments. At $d=d_u=4$ such a scaling function should be augmented by logarithmic corrections, observation of which, however, requires access to few decades of energies and system sizes that is impossible in our analysis. Hence, such corrections in Eq.~\eqref{eq:ADOSscaling} are ignored when $d=4$. First, we suppress the $L$ dependence of ${\mathcal G}$. On the semimetallic side of the QPT, ${\mathcal G}(x) \sim x^{d-1}$, yielding the desired $\rho(E) \sim |E|^{d-1}$ scaling of ADOS. At the QCP ($\delta=0$), ${\mathcal G}(x)$ must be independent of $\delta$, demanding ${\mathcal G}(x) \sim x^{d/z-1}$, yielding $\rho (E) \sim |E|^{d/z-1}$. Therefore, from the scaling of $\rho(E)$ with $E$ at $W=W_c$, one accurately estimates the DSE $z$ at the WSM-to-metal QCP. In the CDM phase, as $\rho(0)$ becomes finite, ${\mathcal G}(x) \sim x^0$, yielding $\rho (0) \sim \delta^{(d-z)\nu}$ with $\beta=(d-z)\nu$ as the order parameter exponent. From the scaling of $\rho(0)$ in the metallic phase, we extract $\beta$ and subsequently the CLE $\nu$ with the predetermined value of $z$. We follow this prescription to determine the universality class of the WSM-to-metal QPT in various $d$.

\emph{Results}.~At first, to test the expected ADOS scaling in clean WSMs where the KPM results display oscillations due to infinitely sharp spectra, we consider weakly disordered systems. Comparing $\rho(E)$ with $|E|^{d-1}$ we observe linear scaling for $d=2$ for $W \ll 1$ and in $d=3, \cdots, 6$ for $W$ far from $W_c$. See Fig.~\ref{fig:clean_2DBCS}(a). In $d=2$, we find BCS-like scaling $\rho(0) \sim \exp[-\lambda/W]$ with $\lambda$ as a fitting parameter; see Fig.~\ref{fig:clean_2DBCS}(b). For $ 3 \leq d \leq 6$, we find that $\rho(0)$ remains almost zero (within numerical accuracy) up to a critical strength of disorder ($W_c$), only beyond which it becomes finite, indicating onset of a CDM through a WSM-to-metal QPT. These features are displayed in $d=3$ [Fig.~\ref{fig:dataanalysis}(a)], $d=4$ [Fig.~\ref{fig:dataanalysis}(e)], $d=5$ [Fig.~\ref{fig:dataanalysis}(i)], and $d=6$ [Fig.~\ref{fig:dataanalysis}(m)]. From the scaling of $\Delta\rho(0)$ with $W$, we find $W_c = 2.65 \pm 0.05$ in $d=3$ [Fig.~\ref{fig:dataanalysis}(b)], $W_c = 3.50 \pm 0.05$ in $d=4$ [Fig.~\ref{fig:dataanalysis}(f)], $W_c = 4.05 \pm 0.05$ in $d=5$ [Fig.~\ref{fig:dataanalysis}(j)], and $W_c = 4.55 \pm 0.05$ in $d=6$ [Fig.~\ref{fig:dataanalysis}(n)]. Error in the determination of $W_c$ stems from the step size in $W$ in our simulations in its close vicinity. A finite $W_c$ in WSMs in $d \geq 3$ is consistent with the predictions from the SCBA.

While determining $W_c$, we fit $\Delta \rho (0)$ with $(W-W_c)^{\beta_{\rm th}}$ deep inside the CDM phase, with the theoretically predicted values of $z=d/2$ and $\nu=1.0$ (in $d=3$) or $0.5$ (in $d=4,5$, and $6$), where it shows excellent agreement with numerical data. Such a curve is extrapolated down to the $\Delta \rho(0)=0$ line to pin $W=W_c$, which we use for the remainder of the analysis. Obtained this way, $W_c$ is slightly higher than the value found directly from the sign change in the raw data of $\Delta \rho(0)$.

Near $W=W_c$, by fitting $\rho(E)$ with $|E|^{d/z-1}$ over a finite range of energy near $E=0$ (marked by vertical dashed lines), we obtain $z = 1.505 \pm 0.049$ in $d=3$ [Fig.~\ref{fig:dataanalysis}(c)], $z = 2.065 \pm 0.069$ in $d=4$ [Fig.~\ref{fig:dataanalysis}(g)], $z = 2.500 \pm 0.156$ in $d=5$ [Fig.~\ref{fig:dataanalysis}(k)], and $z = 2.994 \pm 0.258$ in $d=6$ [Fig.~\ref{fig:dataanalysis}(o)]. The quoted error bars result from fitting error. These findings strongly suggest that in any $d \geq 3$, $z =d/2$ (within numerical accuracy), which we soon show to be an exact result from an RG calculation.

Next we focus on the metallic phase to extract $\beta=(d-z)\nu$ and subsequently $\nu$ from the existing knowledge of $z$ by comparing $\rho(0)$ with $\delta$ therein. To stay away from the quantum critical regime, we discard a few points of $\rho(0)$ for sufficiently small $\delta$ and focus on such a scaling deep inside the CDM phase. We then obtain $\nu= 1.017 \pm 0.034$ in $d=3$ [Fig.~\ref{fig:dataanalysis}(d)], $\nu = 0.563 \pm 0.020$ in $d=4$ [Fig.~\ref{fig:dataanalysis}(h)], $\nu = 0.518 \pm 0.032$ in $d=5$ [Fig.~\ref{fig:dataanalysis}(l)], and $\nu = 0.502 \pm 0.043$ in $d=6$ [Fig.~\ref{fig:dataanalysis}(p)]. These findings strongly suggest that $d=4$ stands as the UCD for the WSM-to-metal QPT, as at and above which the CLE gets locked (within numerical accuracy) to its mean-field or Gaussian value of $\nu =0.5$.

To test the validity of the numerically extracted values of $W_c$, $z$, and $\nu$, next we search for single-parameter scalings across the WSM-to-metal QCP in $d=3, \cdots, 6$. Motivated by the scaling form in Eq.~\eqref{eq:ADOSscaling} and neglecting the $L$ dependence therein, first we compare $\rho(E) \delta^{-\nu(d-z)}$ with $|E|\delta^{-\nu z}$ over wide ranges of energy and disorder. All data collapse onto three curves with the mean values of $W_c$, $z$, and $\nu$ for $d=3$ [Fig.~\ref{fig:datacollapse}(a)], $d=4$ [Fig.~\ref{fig:datacollapse}(c)], $d=5$ [Fig.~\ref{fig:datacollapse}(e)], and $d=6$ [Fig.~\ref{fig:datacollapse}(g)]. The lower and upper branches in the small $|E|\delta^{-\nu z}$ regime stem from the WSM and CDM sides of the QPT, respectively, which meet on a common curve for large $|E|\delta^{-\nu z}$, resulting from the quantum critical regime. To further corroborate the single-parameter scaling, next we concentrate on the CDM phase, and upon setting $E=0$ in Eq.~\eqref{eq:ADOSscaling}, compare $\rho(0) L^{d-z}$ with $L^{1/\nu} \delta$, as the correlation length $\xi \sim |\delta|^{-\nu} \sim L$ (finite) therein. Data obtained in various systems with $L \leq L_{\rm max}$ collapse onto a single curve in $d=3$ [Fig.~\ref{fig:datacollapse}(b)], $d=4$ [Fig.~\ref{fig:datacollapse}(d)], $d=5$ [Fig.~\ref{fig:datacollapse}(f)], and $d=6$ [Fig.~\ref{fig:datacollapse}(h)] with the mean values of $W_c$, $z$, and $\nu$. These two sets of high-quality data collapses strongly endorse the central claim of this work that $d_u=4$ for WSM-to-metal QPT and $z=d/2$ therein for any $d \geq 3$.

\emph{RG analysis}.~The starting point of the RG analysis, addressing the stability of WSMs for weak disorder and their QPT into a CDM at stronger disorder in any $d>2$ within the framework of an $\epsilon_d$ expansion with $\epsilon_d=d-2$, is the imaginary time ($\tau$) replicated action ($\bar{S}$). Performing an average over pointlike charge impurities $V(\vec{r})\Gamma_0$ assuming a Gaussian white noise distribution with zero mean, i.e., $\langle \langle V(\vec{r}) V(\vec{r}')\rangle \rangle= \Delta_0 \delta^{d} (\vec{r}-\vec{r}')$, where $\delta^{d}(x)$ is the $d$-dimensional Dirac delta function, we find 
\begin{eqnarray}~\label{SMeq:replicaS}
\bar{S} &=& \int d^d\vec{r} d\tau \; \bar{\Psi}_\alpha(\tau,\vec{r}) \left(\gamma_0 \partial_0 + v \gamma_j \partial_j \right) \Psi_\alpha(\tau,\vec{r}) \nonumber \\ 
&-& \frac{\Delta_0}{2} \int d^d\vec{r} d\tau d\tau' \left[ \bar{\Psi}_\alpha \gamma_0 \Psi_\alpha\right]_{(\tau,\vec{r})} \;
 \left[ \bar{\Psi}_\beta \gamma_0 \Psi_\beta\right]_{(\tau',\vec{r})}.\;\:\:
\end{eqnarray} 
Here, $\alpha, \beta$ are replica indices, a summation over repeated spatial index $j=1, \cdots, d$ is assumed, $\Gamma_j =i \gamma_0 \gamma_j$ with the Hermitian $\gamma$ matrices satisfying the anticommuting Clifford algebra $\{ \gamma_\mu, \gamma_\nu \}=2 \delta_{\mu \nu}$ (Kronecker delta symbol), and $\bar{\Psi} \equiv \Psi^\dagger \gamma_0$ and $\Psi$ are independent Grassmann variables. The scale invariance of $\bar{S}$ dictates scaling dimensions for the Fermi velocity $[v]=1-z$ and disorder coupling $[\Delta_0]=2 z-d= 2 (z-1)-\epsilon_d$~\cite{goswami2011, Syzranov2015a, RoyDasSarma2016PRB}. Computing the one-loop Feynman diagrams, we arrive at the RG flow equations, captured by the $\beta$-functions $\beta_v=-\Delta_0 \; v \equiv (1-z)v$ and  
\begin{equation}
\beta_{\Delta_0}= \left[ -\epsilon_d + 2 (z-1) \right] \Delta_0 = \left[ -\epsilon_d + 2 \Delta_0 \right] \Delta_0
\end{equation}
where $2 \Lambda^\epsilon \Delta_0/[(4 \pi)^d v^2] \to \Delta_0$ is the dimensionless disorder coupling with $\Lambda$ setting a running (frequency or momentum) scale. See Supplemental Material (SM) for details~\cite{SMUpperCritical}. From $\beta_{\Delta_0}=0$ we find two fixed points: (1) a stable one at $\Delta_0=0$, describing a robust WSM phase for weak disorder in any $d>2$ and (2) an unstable QCP at $\Delta_0=\epsilon_d/2$, where the DSE is $z=1+\epsilon_d/2=d/2$ and $\nu^{-1}=\epsilon_d=d-2$. However, in this scheme both exponents receive higher-order corrections, causing increasing discrepancies with numerical findings in $d=3$~\cite{Syzranov2016, roydassarma2016, Carpentier2016}. Still, within the leading-order $\epsilon_d$ expansion about the lower critical two spatial dimensions, we find $z=d/2$ in any $d$, and $\nu=1$ and $1/2$ in $d=3$ and $d=4$, respectively, which are consistent with our numerical findings. The fact that $\nu=1/2$ (mean-field value) in $d=4$ confirms that it is the UCD for the WSM-to-metal QPT. However, an expansion about the LCD cannot correctly predict the CLE above the UCD, a known fact in critical phenomena. By the same token, any expansion about the UCD is oblivious to the existence of the LCD~\cite{ZinnJustin2002, Herbut2007, Sachdev2011}. The latter approach, tailored for the WSM-to-metal QPT about $d_u=4$ in terms of the $Q$ matrix~\cite{Goswami2017b, Carpentier2016}, correctly predicts $\nu=1/2$ in any $d \geq 4$.

Next, we briefly discuss an alternative RG approach which predicts $z=d/2$ as an exact result in any $d>2$, despite suffering similar limitations while predicting $\nu$. In this case, the imaginary time action is~\cite{Moon2014, Goswami2017b, Slager2018} 
\begin{equation}~\label{eq:action}
S=\int d^d {\vec r} d\tau \left[{\bar \Psi}(\gamma_0\partial_\tau+v\gamma_j\partial_j)\Psi-\varphi_{_0}({\bar \Psi} \gamma_0{ \Psi})\right],
\end{equation}
where $\varphi_{_0}$ is the disorder field, obeying the distribution $\langle \varphi_{_0}({\vec r}) \varphi_{_0}({\vec r}^\prime)\rangle={\Delta}_0/|{\vec r}-{\vec r}^\prime|^{d-m}$ in real space or $\langle \varphi_{_0}({\vec q}) \varphi_{_0}({\vec 0}) \rangle={\bar \Delta}_0/|{\vec q}|^{m}$ in reciprocal space. The limit $m \rightarrow 0$ corresponds to the Gaussian white noise distribution, which we are ultimately interested in. Such a form of the white noise distribution stems from the following representation of the $d$-dimensional Dirac $\delta$ function~\cite{Stein1971}
\allowdisplaybreaks[4]
\begin{equation}
\delta^{(d)}({\vec r}-{\vec r}^\prime)=\lim_{m\rightarrow0}\frac{\Gamma\left(\frac{d-m}{2}\right)}{2^m\pi^{d/2}\Gamma(m/2)} \: \frac{1}{|{\vec r}-{\vec r}^\prime|^{d-m}}.
\end{equation}
The scale invariance of $S$ yields the following scaling dimensions $[v]=1-z$, $[\varphi_{_0}]=z+\eta_{\varphi_{_0}}$, where $\eta_{\varphi_{_0}}$ is the anomalous dimension of the disorder field, and $[\Delta_0]=-\epsilon_m + 2 (z-1)+2 \eta_{\varphi_{_0}}$ with $\epsilon_m=d-m-2$. The non-analytic structure of the disorder field propagator along with a reduced U(1) gauge symmetry that leaves $S$ invariant under $\varphi_{_0} \to \varphi_{_0} + \partial_\tau \lambda(\tau)$, $\bar{\Psi} \to \exp[-i\lambda(\tau)] \bar{\Psi}$, and $\Psi \to \exp[i\lambda(\tau)] \Psi$, where $\lambda(\tau)$ is an arbitrary scalar function of imaginary time, ensures that $\eta_{\varphi_{_0}} \equiv 0$ (manifestation of the Ward identity), which we explicitly verify to the one-loop order in the SM~\cite{SMUpperCritical}. Then the $\beta$-function for the (dimensionless) disorder coupling takes the exact form of $\beta_{\Delta_0}=[-\epsilon_m + 2 (z-1)]\Delta_0$, which yields an exact result of $z=1+\epsilon_m/2 \to d/2$ for Gaussian white noise distribution for any $d>2$. To the leading order in $\epsilon_m$ expansion, the CLE is given by $\nu^{-1}=\epsilon_m=d-2$ when $m \to 0$, an outcome similar to the one from $\epsilon_d$ expansion. However, ${\mathcal O}(\epsilon^2_m)$ corrections to $\nu$ is currently unknown. Besides the exactness of $z=d/2$ in any $d>2$, which is consistent with our numerical findings, the $\epsilon_m$ expansion also predicts a nontrivial fermionic anomalous dimension $\eta_{_\Psi}=3 \epsilon_m/8$ near the WSM-to-metal QCP~\cite{Goswami2017b, Slager2018}. Such a prediction has been verified numerically in $d=3$~\cite{Goswami2017a}, making $\epsilon_m$ expansion reliable over the traditional $\epsilon_d$ one.

\emph{Summary and discussions}.~From extensive numerical (ADOS using KPM) and analytical (SCBA and RG) calculations, here we present compelling evidence that four (two) spatial dimensions stand as the UCD (LCD) for the disorder-driven WSM-to-metal QPT. Consequently, the CLE gets locked to its mean-field value of $\nu=0.5$ for $d \geq 4$, while the DSE is exactly $z=d/2$ (shown from the $\epsilon_m$ expansion). These findings justify the existing Yukawa-type field-theoretic approach to address to the WSM-to-metal QPT in terms of the $Q$ matrix and a controlled $\epsilon_u$ expansion about the upper critical four spatial dimensions with $\epsilon_u=4-d$, which guarantees $\nu=1/2$ for any $d \geq 4$. However, the fact that $\nu \approx 1$ in $d=3$ remains an unresolved issue, which can possibly be settled through a higher-order $\epsilon_m$ expansion. To further establish the mean-field nature of such a QPT in $d \geq 4$, it will be worthwhile to show that the fermionic anomalous dimension $\eta_\Psi \to 0$ therein. When $W_c$ is located from the sign change in raw data of $\Delta \rho(0)$, the values of $\nu$ deviate from $0.5$ in $d \geq 4$, which in general leads to worse data collapses~\cite{SMUpperCritical}. We also believe that these outcomes hold for any arbitrary type of disorder, which is left for future investigations. Unfortunately, here we could not demonstrate the logarithmic violation of scaling at $d=4$ as the tight-binding model for WSMs with finite-range (local) hopping produces a linear energy-momentum relation only over a small segment of the Brillouin zone near $E=0$. SLAC fermions, featuring Weyl dispersion over the entire reciprocal space~\cite{Drell1976}, on the other hand, can be the ideal platform to demonstrate logarithmic violation of scaling at $d=4$. However, the nonlocal hopping amplitudes in the construction of SLAC fermions make KPM, ideally suited for sparse matrices, numerically expensive. We, thus, leave this issue for a future study.

Altogether, results presented in this work give us the luxury to close the discussion with a fascinating comparative anecdote on Dirac or Weyl semimetals, in the presence of (a) Hubbard-like local or short-range interactions, sourcing inelastic scattering and responsible for dynamic mass generation via spontaneous symmetry breaking or (b) disorder, causing elastic scattering, which can give birth to CDM. In the former situation, $d_\ell=1$ and $d_u=3$. Therefore, QPTs in Dirac-Weyl materials, triggered by inelastic or elastic scattering, can always be studied in terms of perturbative field-theoretic approaches about both the LCD and UCD. For example, the phenomenon of dynamic mass generation in Dirac-Weyl materials can be studied in terms of the Gross-Neveu~\cite{Gross1974} and Nambu-Jona-Lasinio~\cite{Nambu1961} models about $d_\ell=1$ or in terms of the Gross-Neveu-Yukawa field theory about $d_u=4$~\cite{ZinnJustin2002}. By the same token, the Dirac-Weyl semimetal-to-metal QPTs can be studied (at least qualitatively) within the frameworks of $\epsilon$ expansions about the lower critical two or upper critical four spatial dimensions.

\emph{Acknowledgments}.~B.R.\ was supported by NSF CAREER Grant No.\ DMR-2238679 and thanks Vladimir Juri\v{c}i\'c for useful discussion. Portions of this research were conducted on Lehigh University's Research Computing infrastructure partially supported by NSF Award No.~2019035.

\bibliography{References_Upper_Critical_Dimension}

@book{ZinnJustin2002,
  title = "{Quantum Field Theory and Critical Phenomena}",
  ISBN = {9780198509233},
  url = {http://dx.doi.org/10.1093/acprof:oso/9780198509233.001.0001},
  DOI = {10.1093/acprof:oso/9780198509233.001.0001},
  publisher = {Oxford University Press, Oxford, UK},
  author = {Zinn-Justin,  Jean},
  year = {2002},
  month = {June} 
}

@book{Herbut2007,
  title = "{A Modern Approach to Critical Phenomena}",
  ISBN = {9780511755521},
  url = {http://dx.doi.org/10.1017/CBO9780511755521},
  DOI = {10.1017/cbo9780511755521},
  publisher = {Cambridge University Press, Cambridge, UK},
  author = {Herbut, Igor F.},
  year = {2007},
  month = {Jan} 
}

@book{Sachdev2011,
  title = "{Quantum Phase Transitions}",
  ISBN = {9780511973765},
  url = {http://dx.doi.org/10.1017/CBO9780511973765},
  DOI = {10.1017/cbo9780511973765},
  publisher = {Cambridge University Press, Cambridge, UK},
  author = {Sachdev,  Subir},
  year = {2011},
  month = {Apr} 
}

@book{Ping2006,
  title = "{Introduction to Wave Scattering,  Localization and Mesoscopic Phenomena}",
  ISBN = {9783540291565},
  ISSN = {2196-2812},
  url = {http://dx.doi.org/10.1007/3-540-29156-3},
  DOI = {10.1007/3-540-29156-3},
  journal = {Springer Series in MATERIALS SCIENCE},
  publisher = {Springer Berlin Heidelberg},
  author = {Ping,  Sheng},
  year = {2006}
}

@book{Stein1971,
  title = "{Singular Integrals and Differentiability Properties of Functions}",
  ISBN = {9781400883882},
  url = {http://dx.doi.org/10.1515/9781400883882},
  DOI = {10.1515/9781400883882},
  publisher = {Princeton University Press},
  author = {Stein,  Elias M.},
  year = {1971},
  month = {Dec} 
}

@article{Anderson1958,
  title = "{Absence of Diffusion in Certain Random Lattices}",
  volume = {109},
  ISSN = {0031-899X},
  url = {http://dx.doi.org/10.1103/PhysRev.109.1492},
  DOI = {10.1103/physrev.109.1492},
  number = {5},
  journal = {Phys. Rev.},
  publisher = {American Physical Society (APS)},
  author = {Anderson,  P. W.},
  year = {1958},
  month = {Mar},
  pages = {1492}
}

@article{Abrahams1979,
  title = "{Scaling Theory of Localization: Absence of Quantum Diffusion in Two Dimensions}",
  author = {Abrahams, E. and Anderson, P. W. and Licciardello, D. C. and Ramakrishnan, T. V.},
  journal = {Phys. Rev. Lett.},
  volume = {42},
  issue = {10},
  pages = {673},
  numpages = {0},
  year = {1979},
  month = {Mar},
  publisher = {American Physical Society},
  doi = {10.1103/PhysRevLett.42.673},
  url = {https://link.aps.org/doi/10.1103/PhysRevLett.42.673}
}

@article{Wegner1976,
  title = "{Electrons in disordered systems. Scaling near the mobility edge}",
  volume = {25},
  ISSN = {1434-6036},
  url = {http://dx.doi.org/10.1007/BF01315248},
  DOI = {10.1007/bf01315248},
  number = {4},
  journal = {Z. Phys. B: Condens. Matter},
  publisher = {Springer Science and Business Media LLC},
  author = {Wegner,  Franz J.},
  year = {1976},
  month = {Dec},
  pages = {327}
}

@article{Wegner1980,
  title = "{Inverse participation ratio in $2+\epsilon$ dimensions}",
  volume = {36},
  ISSN = {1434-6036},
  url = {http://dx.doi.org/10.1007/BF01325284},
  DOI = {10.1007/bf01325284},
  number = {3},
  journal = {Z. Phys. B: Condens. Matter},
  publisher = {Springer Science and Business Media LLC},
  author = {Wegner,  F.},
  year = {1980},
  month = {Sept},
  pages = {209}
}

@article{Vollhardt1982,
  title = "{Scaling Equations from a Self-Consistent Theory of Anderson Localization}",
  author = {Vollhardt, D. and W\"olfle, P.},
  journal = {Phys. Rev. Lett.},
  volume = {48},
  issue = {10},
  pages = {699},
  numpages = {0},
  year = {1982},
  month = {Mar},
  publisher = {American Physical Society},
  doi = {10.1103/PhysRevLett.48.699},
  url = {https://link.aps.org/doi/10.1103/PhysRevLett.48.699}
}

@article{Schreiber1996,
  title = "{Dimensionality Dependence of the Metal-Insulator Transition in the Anderson Model of Localization}",
  author = {Schreiber, Michael and Grussbach, Heiko},
  journal = {Phys. Rev. Lett.},
  volume = {76},
  issue = {10},
  pages = {1687},
  numpages = {0},
  year = {1996},
  month = {Mar},
  publisher = {American Physical Society},
  doi = {10.1103/PhysRevLett.76.1687},
  url = {https://link.aps.org/doi/10.1103/PhysRevLett.76.1687}
}

@article{Lee1985,
  title = {Disordered electronic systems},
  author = {Lee, Patrick A. and Ramakrishnan, T. V.},
  journal = {Rev. Mod. Phys.},
  volume = {57},
  issue = {2},
  pages = {287},
  numpages = {0},
  year = {1985},
  month = {Apr},
  publisher = {American Physical Society},
  doi = {10.1103/RevModPhys.57.287},
  url = {https://link.aps.org/doi/10.1103/RevModPhys.57.287}
}

@article{Harris1981,
  title = "{Mean-field theory and $\ensuremath{\epsilon}$ expansion for Anderson localization}",
  author = {Harris, A. B. and Lubensky, T. C.},
  journal = {Phys. Rev. B},
  volume = {23},
  issue = {6},
  pages = {2640},
  numpages = {0},
  year = {1981},
  month = {Mar},
  publisher = {American Physical Society},
  doi = {10.1103/PhysRevB.23.2640},
  url = {https://link.aps.org/doi/10.1103/PhysRevB.23.2640}
}

@article{Mirlin1994,
  title = "{Distribution of local densities of states, order parameter function, and critical behavior near the Anderson transition}",
  author = {Mirlin, Alexander D. and Fyodorov, Yan V.},
  journal = {Phys. Rev. Lett.},
  volume = {72},
  issue = {4},
  pages = {526},
  numpages = {0},
  year = {1994},
  month = {Jan},
  publisher = {American Physical Society},
  doi = {10.1103/PhysRevLett.72.526},
  url = {https://link.aps.org/doi/10.1103/PhysRevLett.72.526}
}

@article{Castellani1986,
  title = "{On the upper critical dimension in Anderson localisation}",
  volume = {19},
  ISSN = {1361-6447},
  url = {http://dx.doi.org/10.1088/0305-4470/19/17/009},
  DOI = {10.1088/0305-4470/19/17/009},
  number = {17},
  journal = {J. Phys. A: Math. Gen.},
  publisher = {IOP Publishing},
  author = {Castellani,  C and Castro,  C Di and Peliti,  L},
  year = {1986},
  month = Dec,
  pages = {L1099}
}

@article{Garcia2007,
  title = "{Dimensional dependence of the metal-insulator transition}",
  author = {Garc\'{\i}a-Garc\'{\i}a, Antonio M. and Cuevas, Emilio},
  journal = {Phys. Rev. B},
  volume = {75},
  issue = {17},
  pages = {174203},
  numpages = {8},
  year = {2007},
  month = {May},
  publisher = {American Physical Society},
  doi = {10.1103/PhysRevB.75.174203},
  url = {https://link.aps.org/doi/10.1103/PhysRevB.75.174203}
}

@article{Mard2017,
  title = "{Strong-disorder approach for the Anderson localization transition}",
  author = {Mard, H. Javan and Hoyos, Jos\'e A. and Miranda, E. and Dobrosavljevi\ifmmode \acute{c}\else \'{c}\fi{}, V.},
  journal = {Phys. Rev. B},
  volume = {96},
  issue = {4},
  pages = {045143},
  numpages = {5},
  year = {2017},
  month = {Jul},
  publisher = {American Physical Society},
  doi = {10.1103/PhysRevB.96.045143},
  url = {https://link.aps.org/doi/10.1103/PhysRevB.96.045143}
}

@article{Fradkin1986,
  title = "{Critical behavior of disordered degenerate semiconductors. II. Spectrum and transport properties in mean-field theory}",
  author = {Fradkin, Eduardo},
  journal = {Phys. Rev. B},
  volume = {33},
  issue = {5},
  pages = {3263},
  numpages = {0},
  year = {1986},
  month = {Mar},
  publisher = {American Physical Society},
  doi = {10.1103/PhysRevB.33.3263},
  url = {https://link.aps.org/doi/10.1103/PhysRevB.33.3263}
}

@article{Murakami2009,
  title = "{Effects of disorder in three-dimensional ${Z}_{2}$ quantum spin Hall systems}",
  author = {Shindou, Ryuichi and Murakami, Shuichi},
  journal = {Phys. Rev. B},
  volume = {79},
  issue = {4},
  pages = {045321},
  numpages = {29},
  year = {2009},
  month = {Jan},
  publisher = {American Physical Society},
  doi = {10.1103/PhysRevB.79.045321},
  url = {https://link.aps.org/doi/10.1103/PhysRevB.79.045321}
}

@article{goswami2011,
  title = "{Quantum Criticality between Topological and Band Insulators in $3+1$ Dimensions}",
  author = {Goswami, P. and Chakravarty, S.},
  journal = {Phys. Rev. Lett.},
  volume = {107},
  issue = {19},
  pages = {196803},
  numpages = {5},
  year = {2011},
  month = {Nov},
  publisher = {American Physical Society},
  doi = {10.1103/PhysRevLett.107.196803}
}

@article{Imura2013,
  title = "{Disordered Weak and Strong Topological Insulators}",
  author = {Kobayashi, Koji and Ohtsuki, Tomi and Imura, Ken-Ichiro},
  journal = {Phys. Rev. Lett.},
  volume = {110},
  issue = {23},
  pages = {236803},
  numpages = {5},
  year = {2013},
  month = {Jun},
  publisher = {American Physical Society},
  doi = {10.1103/PhysRevLett.110.236803},
  url = {https://link.aps.org/doi/10.1103/PhysRevLett.110.236803}
}

@article{Herbut2014,
  title = {Density of States Scaling at the Semimetal to Metal Transition in Three Dimensional Topological Insulators},
  author = {Kobayashi, Koji and Ohtsuki, Tomi and Imura, Ken-Ichiro and Herbut, Igor F.},
  journal = {Phys. Rev. Lett.},
  volume = {112},
  issue = {1},
  pages = {016402},
  numpages = {5},
  year = {2014},
  month = {Jan},
  publisher = {American Physical Society},
  doi = {10.1103/PhysRevLett.112.016402},
  url = {https://link.aps.org/doi/10.1103/PhysRevLett.112.016402}
}

@article{roydassarma2014,
  title = "{Diffusive quantum criticality in three-dimensional disordered Dirac semimetals}",
  author = {Roy, Bitan and Das Sarma, S.},
  journal = {Phys. Rev. B},
  volume = {90},
  issue = {24},
  pages = {241112},
  numpages = {5},
  year = {2014},
  month = {Dec},
  publisher = {American Physical Society},
  doi = {10.1103/PhysRevB.90.241112},
  url = {https://link.aps.org/doi/10.1103/PhysRevB.90.241112}
}

@misc{Moon2014,
  author = {Moon,  Eun-Gook and Kim,  Yong Baek},
  title = "{Non-Fermi Liquid in Dirac Semi-metals}",
  Eprint= {arXiv:1409.0573 (2014)}
}

@article{Syzranov2015a,
  title = "{Critical Transport in Weakly Disordered Semiconductors and Semimetals}",
  author = {Syzranov, S. V. and Radzihovsky, L. and Gurarie, V.},
  journal = {Phys. Rev. Lett.},
  volume = {114},
  issue = {16},
  pages = {166601},
  numpages = {5},
  year = {2015},
  month = {Apr},
  publisher = {American Physical Society},
  doi = {10.1103/PhysRevLett.114.166601},
  url = {https://link.aps.org/doi/10.1103/PhysRevLett.114.166601}
}

@article{Syzranov2015b,
  title = {Unconventional localization transition in high dimensions},
  author = {Syzranov, S. V. and Gurarie, V. and Radzihovsky, L.},
  journal = {Phys. Rev. B},
  volume = {91},
  issue = {3},
  pages = {035133},
  numpages = {15},
  year = {2015},
  month = {Jan},
  publisher = {American Physical Society},
  doi = {10.1103/PhysRevB.91.035133},
  url = {https://link.aps.org/doi/10.1103/PhysRevB.91.035133}
}

@article{Altland2015,
  title = "{Effective Field Theory of the Disordered Weyl Semimetal}",
  author = {Altland, Alexander and Bagrets, Dmitry},
  journal = {Phys. Rev. Lett.},
  volume = {114},
  issue = {25},
  pages = {257201},
  numpages = {5},
  year = {2015},
  month = {Jun},
  publisher = {American Physical Society},
  doi = {10.1103/PhysRevLett.114.257201},
  url = {https://link.aps.org/doi/10.1103/PhysRevLett.114.257201}
}

@article{pix2015,
  title = "{Anderson Localization and the Quantum Phase Diagram of Three Dimensional Disordered Dirac Semimetals}",
  author = {Pixley, J. H. and Goswami, Pallab and Das Sarma, S.},
  journal = {Phys. Rev. Lett.},
  volume = {115},
  issue = {7},
  pages = {076601},
  numpages = {5},
  year = {2015},
  month = {Aug},
  publisher = {American Physical Society},
  doi = {10.1103/PhysRevLett.115.076601},
  url = {https://link.aps.org/doi/10.1103/PhysRevLett.115.076601}
}

@article{Syzranov2016,
  title = "{Critical exponents at the unconventional disorder-driven transition in a Weyl semimetal}",
  author = {Syzranov, S. V. and Ostrovsky, P. M. and Gurarie, V. and Radzihovsky, L.},
  journal = {Phys. Rev. B},
  volume = {93},
  issue = {15},
  pages = {155113},
  numpages = {10},
  year = {2016},
  month = {Apr},
  publisher = {American Physical Society},
  doi = {10.1103/PhysRevB.93.155113},
  url = {https://link.aps.org/doi/10.1103/PhysRevB.93.155113}
}

@article{Pix2016a,
  title = "{Disorder-driven itinerant quantum criticality of three-dimensional massless Dirac fermions}",
  author = {Pixley, J. H. and Goswami, Pallab and Das Sarma, S.},
  journal = {Phys. Rev. B},
  volume = {93},
  issue = {8},
  pages = {085103},
  numpages = {16},
  year = {2016},
  month = {Feb},
  publisher = {American Physical Society},
  doi = {10.1103/PhysRevB.93.085103},
  url = {https://link.aps.org/doi/10.1103/PhysRevB.93.085103}
}

@article{roydassarma2016,
  title = "{Erratum: Diffusive quantum criticality in three-dimensional disordered Dirac semimetals [Phys.\ Rev.\ B {\bf 90}, 241112(R) (2014)]}",
  author = {Roy, Bitan and Das Sarma, S.},
  journal = {Phys. Rev. B},
  volume = {93},
  issue = {11},
  pages = {119911},
  numpages = {3},
  year = {2016},
  month = {Mar},
  publisher = {American Physical Society},
  doi = {10.1103/PhysRevB.93.119911},
  url = {https://link.aps.org/doi/10.1103/PhysRevB.93.119911}
}

@article{bera2016,
  title = "{Dirty Weyl semimetals: Stability, phase transition, and quantum criticality}",
  author = {Bera, Soumya and Sau, Jay D. and Roy, Bitan},
  journal = {Phys. Rev. B},
  volume = {93},
  issue = {20},
  pages = {201302},
  numpages = {5},
  year = {2016},
  month = {May},
  publisher = {American Physical Society},
  doi = {10.1103/PhysRevB.93.201302},
  url = {https://link.aps.org/doi/10.1103/PhysRevB.93.201302}
}

@article{Gorbar2016,
  title = "{Origin of dissipative Fermi arc transport in Weyl semimetals}",
  author = {Gorbar, E. V. and Miransky, V. A. and Shovkovy, I. A. and Sukhachov, P. O.},
  journal = {Phys. Rev. B},
  volume = {93},
  issue = {23},
  pages = {235127},
  numpages = {17},
  year = {2016},
  month = {Jun},
  publisher = {American Physical Society},
  doi = {10.1103/PhysRevB.93.235127},
  url = {https://link.aps.org/doi/10.1103/PhysRevB.93.235127}
}

@article{RoyJuricicDasSarma2016,
  title = "{Universal optical conductivity of a disordered Weyl semimetal}",
  volume = {6},
  pages = {32446},
  ISSN = {2045-2322},
  url = {http://dx.doi.org/10.1038/srep32446},
  DOI = {10.1038/srep32446},
  number = {1},
  journal = {Sci. Rep.},
  publisher = {Springer Science and Business Media LLC},
  author = {Roy,  Bitan and Juri\ifmmode \check{c}\else \v{c}\fi{}i\ifmmode \acute{c}\else \'{c}\fi{},  Vladimir and Das Sarma,  Sankar},
  year = {2016},
  month = {aug} 
}

@article{RoyDasSarma2016PRB,
  title = "{Quantum phases of interacting electrons in three-dimensional dirty Dirac semimetals}",
  author = {Roy, Bitan and Das Sarma, Sankar},
  journal = {Phys. Rev. B},
  volume = {94},
  issue = {11},
  pages = {115137},
  numpages = {25},
  year = {2016},
  month = {Sep},
  publisher = {American Physical Society},
  doi = {10.1103/PhysRevB.94.115137},
  url = {https://link.aps.org/doi/10.1103/PhysRevB.94.115137}
}

@article{Carpentier2016,
  title = "{On the disorder-driven quantum transition in three-dimensional relativistic metals}",
  author = {Louvet, T. and Carpentier, D. and Fedorenko, A. A.},
  journal = {Phys. Rev. B},
  volume = {94},
  issue = {22},
  pages = {220201},
  numpages = {5},
  year = {2016},
  month = {Dec},
  publisher = {American Physical Society},
  doi = {10.1103/PhysRevB.94.220201},
  url = {https://link.aps.org/doi/10.1103/PhysRevB.94.220201}
}

@article{Carpentier2017,
  title = "{New quantum transition in Weyl semimetals with correlated disorder}",
  author = {Louvet, T. and Carpentier, D. and Fedorenko, A. A.},
  journal = {Phys. Rev. B},
  volume = {95},
  issue = {1},
  pages = {014204},
  numpages = {6},
  year = {2017},
  month = {Jan},
  publisher = {American Physical Society},
  doi = {10.1103/PhysRevB.95.014204},
  url = {https://link.aps.org/doi/10.1103/PhysRevB.95.014204}
}

@article{Justin2017,
  title = "{Quantum phases of disordered three-dimensional Majorana-Weyl fermions}",
  author = {Wilson, Justin H. and Pixley, J. H. and Goswami, Pallab and Das Sarma, S.},
  journal = {Phys. Rev. B},
  volume = {95},
  issue = {15},
  pages = {155122},
  numpages = {22},
  year = {2017},
  month = {Apr},
  publisher = {American Physical Society},
  doi = {10.1103/PhysRevB.95.155122},
  url = {https://link.aps.org/doi/10.1103/PhysRevB.95.155122}
}

@article{Goswami2017a,
  title = "{Single-particle excitations in disordered Weyl fluids}",
  author = {Pixley, J. H. and Chou, Yang-Zhi and Goswami, Pallab and Huse, David A. and Nandkishore, Rahul and Radzihovsky, Leo and Das Sarma, S.},
  journal = {Phys. Rev. B},
  volume = {95},
  issue = {23},
  pages = {235101},
  numpages = {10},
  year = {2017},
  month = {Jun},
  publisher = {American Physical Society},
  doi = {10.1103/PhysRevB.95.235101},
  url = {https://link.aps.org/doi/10.1103/PhysRevB.95.235101}
}

@article{Goswami2017b,
  title = "{Superuniversality of topological quantum phase transition and global phase diagram of dirty topological systems in three dimensions}",
  author = {Goswami, Pallab and Chakravarty, Sudip},
  journal = {Phys. Rev. B},
  volume = {95},
  issue = {7},
  pages = {075131},
  numpages = {21},
  year = {2017},
  month = {Feb},
  publisher = {American Physical Society},
  doi = {10.1103/PhysRevB.95.075131},
  url = {https://link.aps.org/doi/10.1103/PhysRevB.95.075131}
}

@article{Alavirad2017,
  title = "{Global Phase Diagram of a Three-Dimensional Dirty Topological Superconductor}",
  author = {Roy, Bitan and Alavirad, Yahya and Sau, Jay D.},
  journal = {Phys. Rev. Lett.},
  volume = {118},
  issue = {22},
  pages = {227002},
  numpages = {5},
  year = {2017},
  month = {Jun},
  publisher = {American Physical Society},
  doi = {10.1103/PhysRevLett.118.227002},
  url = {https://link.aps.org/doi/10.1103/PhysRevLett.118.227002}
}

@article{Slager2017,
  title = "{Dissolution of topological Fermi arcs in a dirty Weyl semimetal}",
  author = {Slager, Robert-Jan and Juri\ifmmode \check{c}\else \v{c}\fi{}i\ifmmode \acute{c}\else \'{c}\fi{}, Vladimir and Roy, Bitan},
  journal = {Phys. Rev. B},
  volume = {96},
  issue = {20},
  pages = {201401},
  numpages = {6},
  year = {2017},
  month = {Nov},
  publisher = {American Physical Society},
  doi = {10.1103/PhysRevB.96.201401},
  url = {https://link.aps.org/doi/10.1103/PhysRevB.96.201401}
}

@article{WilsonRefael2018,
  title = "{Do the surface Fermi arcs in Weyl semimetals survive disorder?}",
  author = {Wilson, Justin H. and Pixley, J. H. and Huse, David A. and Refael, Gil and Das Sarma, S.},
  journal = {Phys. Rev. B},
  volume = {97},
  issue = {23},
  pages = {235108},
  numpages = {10},
  year = {2018},
  month = {Jun},
  publisher = {American Physical Society},
  doi = {10.1103/PhysRevB.97.235108},
  url = {https://link.aps.org/doi/10.1103/PhysRevB.97.235108}
}

@article{Slager2018,
  title = "{Global Phase Diagram of a Dirty Weyl Liquid and Emergent Superuniversality}",
  author = {Roy, Bitan and Slager, Robert-Jan and Juri\ifmmode \check{c}\else \v{c}\fi{}i\ifmmode \acute{c}\else \'{c}\fi{}, Vladimir},
  journal = {Phys. Rev. X},
  volume = {8},
  issue = {3},
  pages = {031076},
  numpages = {40},
  year = {2018},
  month = {Sep},
  publisher = {American Physical Society},
  doi = {10.1103/PhysRevX.8.031076},
  url = {https://link.aps.org/doi/10.1103/PhysRevX.8.031076}
}

@article{Carpentier2018,
  title = "{Disorder-Driven Quantum Transition in Relativistic Semimetals: Functional Renormalization via the Porous Medium Equation}",
  author = {Balog, Ivan and Carpentier, David and Fedorenko, Andrei A.},
  journal = {Phys. Rev. Lett.},
  volume = {121},
  issue = {16},
  pages = {166402},
  numpages = {6},
  year = {2018},
  month = {Oct},
  publisher = {American Physical Society},
  doi = {10.1103/PhysRevLett.121.166402},
  url = {https://link.aps.org/doi/10.1103/PhysRevLett.121.166402}
}

@article{Carpentier2019,
  title = "{Multifractality at the Weyl-semimetal--diffusive-metal transition for generic disorder}",
  author = {Brillaux, Eric and Carpentier, David and Fedorenko, Andrei A.},
  journal = {Phys. Rev. B},
  volume = {100},
  issue = {13},
  pages = {134204},
  numpages = {11},
  year = {2019},
  month = {Oct},
  publisher = {American Physical Society},
  doi = {10.1103/PhysRevB.100.134204},
  url = {https://link.aps.org/doi/10.1103/PhysRevB.100.134204}
}

@article{Szabo2020,
  title = "{Dirty higher-order Dirac semimetal: Quantum criticality and bulk-boundary correspondence}",
  author = {Szab\'o, Andr\'as L. and Roy, Bitan},
  journal = {Phys. Rev. Res.},
  volume = {2},
  issue = {4},
  pages = {043197},
  numpages = {11},
  year = {2020},
  month = {Nov},
  publisher = {American Physical Society},
  doi = {10.1103/PhysRevResearch.2.043197},
  url = {https://link.aps.org/doi/10.1103/PhysRevResearch.2.043197}
}

@article{Gruzberg2024,
  title = "{Surface quantum critical phenomena in disordered Dirac semimetals}",
  author = {Brillaux, Eric and Fedorenko, Andrei A. and Gruzberg, Ilya A.},
  journal = {Phys. Rev. B},
  volume = {109},
  issue = {17},
  pages = {174204},
  numpages = {18},
  year = {2024},
  month = {May},
  publisher = {American Physical Society},
  doi = {10.1103/PhysRevB.109.174204},
  url = {https://link.aps.org/doi/10.1103/PhysRevB.109.174204}
}

@misc{Tyner2025,
  author = {Tyner,  Alexander C. and Juri\ifmmode \check{c}\else \v{c}\fi{}i\ifmmode \acute{c}\else \'{c}\fi{},  Vladimir and Roy,  Bitan},
  title = "{Two-dimensional Disordered Projected Branes: Stability and Quantum Criticality via Dimensional Reduction}",
  Eprint= {arXiv:2507.23780 (2025)}
}

@article{Huse2014,
  title = "{Rare region effects dominate weakly disordered three-dimensional Dirac points}",
  author = {Nandkishore, Rahul and Huse, David A. and Sondhi, S. L.},
  journal = {Phys. Rev. B},
  volume = {89},
  issue = {24},
  pages = {245110},
  numpages = {12},
  year = {2014},
  month = {Jun},
  publisher = {American Physical Society},
  doi = {10.1103/PhysRevB.89.245110},
  url = {https://link.aps.org/doi/10.1103/PhysRevB.89.245110}
}

@article{Huse2016a,
  title = "{Rare-Region-Induced Avoided Quantum Criticality in Disordered Three-Dimensional Dirac and Weyl Semimetals}",
  author = {Pixley, J. H. and Huse, David A. and Das Sarma, S.},
  journal = {Phys. Rev. X},
  volume = {6},
  issue = {2},
  pages = {021042},
  numpages = {20},
  year = {2016},
  month = {Jun},
  publisher = {American Physical Society},
  doi = {10.1103/PhysRevX.6.021042},
  url = {https://link.aps.org/doi/10.1103/PhysRevX.6.021042}
}

@article{Huse2016b,
  title = "{Uncovering the hidden quantum critical point in disordered massless Dirac and Weyl semimetals}",
  author = {Pixley, J. H. and Huse, David A. and Das Sarma, S.},
  journal = {Phys. Rev. B},
  volume = {94},
  issue = {12},
  pages = {121107},
  numpages = {5},
  year = {2016},
  month = {Sep},
  publisher = {American Physical Society},
  doi = {10.1103/PhysRevB.94.121107},
  url = {https://link.aps.org/doi/10.1103/PhysRevB.94.121107}
}

@article{buch2018,
  title = "{Vanishing Density of States in Weakly Disordered Weyl Semimetals}",
  author = {Buchhold, Michael and Diehl, Sebastian and Altland, Alexander},
  journal = {Phys. Rev. Lett.},
  volume = {121},
  issue = {21},
  pages = {215301},
  numpages = {5},
  year = {2018},
  month = {Nov},
  publisher = {American Physical Society},
  doi = {10.1103/PhysRevLett.121.215301},
  url = {https://link.aps.org/doi/10.1103/PhysRevLett.121.215301}
}

@article{JustinWilson2020,
  title = "{Avoided quantum criticality in exact numerical simulations of a single disordered Weyl cone}",
  author = {Wilson, Justin H. and Huse, David A. and Das Sarma, S. and Pixley, J. H.},
  journal = {Phys. Rev. B},
  volume = {102},
  issue = {10},
  pages = {100201},
  numpages = {6},
  year = {2020},
  month = {Sep},
  publisher = {American Physical Society},
  doi = {10.1103/PhysRevB.102.100201},
  url = {https://link.aps.org/doi/10.1103/PhysRevB.102.100201}
}

@article{JustinWilson2024,
  title = "{Direct topological insulator transitions in three dimensions are destabilized by nonperturbative effects of disorder}",
  author = {Fu, Yixing and Wilson, Justin H. and Huse, David A. and Pixley, J. H.},
  journal = {Phys. Rev. B},
  volume = {109},
  issue = {20},
  pages = {205106},
  numpages = {11},
  year = {2024},
  month = {May},
  publisher = {American Physical Society},
  doi = {10.1103/PhysRevB.109.205106},
  url = {https://link.aps.org/doi/10.1103/PhysRevB.109.205106}
}

@article{PixleyWilsonReview2021,
  title = "{Rare regions and avoided quantum criticality in disordered Weyl semimetals and superconductor}",
  volume = {435},
  ISSN = {0003-4916},
  url = {http://dx.doi.org/10.1016/j.aop.2021.168455},
  DOI = {10.1016/j.aop.2021.168455},
  journal = {Ann. Phys. (Amsterdam)},
  publisher = {Elsevier BV},
  author = {Pixley,  J.H. and Wilson,  Justin H.},
  year = {2021},
  month = dec,
  pages = {168455}
}

@article{Weisse2006,
  title = "{The kernel polynomial method}",
  author = {Wei\ss{}e, Alexander and Wellein, Gerhard and Alvermann, Andreas and Fehske, Holger},
  journal = {Rev. Mod. Phys.},
  volume = {78},
  issue = {1},
  pages = {275},
  numpages = {0},
  year = {2006},
  month = {Mar},
  publisher = {American Physical Society},
  doi = {10.1103/RevModPhys.78.275},
  url = {https://link.aps.org/doi/10.1103/RevModPhys.78.275}
}

@article{Groth2014,
  doi = {10.1088/1367-2630/16/6/063065},
  url = {https://doi.org/10.1088/1367-2630/16/6/063065},
  year = {2014},
  month = jun,
  publisher = {{IOP} Publishing},
  volume = {16},
  number = {6},
  pages = {063065},
  author = {Christoph W Groth and Michael Wimmer and Anton R Akhmerov and Xavier Waintal},
  title = "{Kwant: a software package for quantum transport}",
  journal = {New J. Phys.}
}

@article{SMUpperCritical,
  journal = {See Supplemental Material at XXX-XXXX for the details of RG calculations and additional numerical analyses},
  volume = {},
  issue = {},
  pages = {},
  numpages = {},
  year = {},
  month = {},
  doi = {}
}

@article{Drell1976,
  title = "{Strong-coupling field theories. II. Fermions and gauge fields on a lattice}",
  author = {Drell, Sidney D. and Weinstein, Marvin and Yankielowicz, Shimon},
  journal = {Phys. Rev. D},
  volume = {14},
  issue = {6},
  pages = {1627--1647},
  numpages = {0},
  year = {1976},
  month = {Sep},
  publisher = {American Physical Society},
  doi = {10.1103/PhysRevD.14.1627},
  url = {https://link.aps.org/doi/10.1103/PhysRevD.14.1627}
}

@article{Gross1974,
  title = "{Dynamical symmetry breaking in asymptotically free field theories}",
  author = {Gross, David J. and Neveu, Andr\'e},
  journal = {Phys. Rev. D},
  volume = {10},
  issue = {10},
  pages = {3235},
  numpages = {0},
  year = {1974},
  month = {Nov},
  publisher = {American Physical Society},
  doi = {10.1103/PhysRevD.10.3235},
  url = {https://link.aps.org/doi/10.1103/PhysRevD.10.3235}
}

@article{Nambu1961,
  title = "{Dynamical Model of Elementary Particles Based on an Analogy with Superconductivity. I}",
  volume = {122},
  ISSN = {0031-899X},
  url = {http://dx.doi.org/10.1103/PhysRev.122.345},
  DOI = {10.1103/physrev.122.345},
  number = {1},
  journal = {Physical Review},
  publisher = {American Physical Society (APS)},
  author = {Nambu,  Y. and Jona-Lasinio,  G.},
  year = {1961},
  month = Apr,
  pages = {345}
}

\end{document}